\documentstyle[aaspp4]{article}

\slugcomment{Ap.J.Letters, in press}

\lefthead{Fujita et al.}
\righthead{Emission from Isolated Black Holes and MACHOs}

\begin{document}

\title{Emission from Isolated Black Holes and MACHOs\\
 Accreting from the Interstellar Medium}

\author{Yutaka Fujita and Susumu Inoue}

\affil{Department of Physics, Tokyo Metropolitan University,\\ 
 Minami-Ohsawa 1-1, Hachioji, Tokyo 192-03, Japan}
\authoremail{yfujita@phys.metro-u.ac.jp}

\author{Takashi Nakamura}

\affil{Yukawa Institute for Theoretical Physics,\\
 Kyoto University, Sakyo-ku, Kyoto 606-01, Japan}

\author{Tadahiro Manmoto and Kenji E. Nakamura}

\affil{Department of Astronomy, Faculty of Science,\\
 Kyoto University, Sakyo-ku, Kyoto 606-01, Japan}

\begin{abstract}
Adopting the advection dominated accretion flow (ADAF) model,
 we have re-examined the radiation emitted by isolated stellar-mass black holes
 accreting from the interstellar medium of our Galaxy.
Two distinct types of black holes are given consideration,
 1) remnants of conventional massive star evolution (RBHs, mass $\sim 10 \rm M_{\sun}$), and
 2) low-mass objects which possibly constitutes the dark halo
 and cause the observed microlensing events (MACHOBHs, mass $\sim 0.5 \rm M_{\sun}$).
In contrast to previous studies incorporating spherical accretion models,
 the characteristic hard X-ray emission from ADAFs
 serves as an important observational signature.
Standard estimates for the present density of RBHs
 suggest that a fair number of objects,
 either in nearby molecular clouds (such as Orion)
 or in the general Galactic disk, may be detected by near future X-ray instruments.
Their observation may also be feasible at lower frequencies,
 particularly in the infrared and optical bands.
On the other hand,
 the smaller mass and larger velocity on average of MACHOBHs compared to RBHs
 make them unlikely to be observable in the foreseeable future.

\end{abstract}


\section{Introduction}

Massive stars are believed to end their lives undergoing gravitational collapse,
 with the more massive ones leaving behind black holes.
The existence of such stellar-mass black holes in Galactic X-ray binary systems,
 radiating through gas accretion from their companion stars,
 has achieved strong observational support
 (e.g. \markcite{tl1996}Tanaka and Lewin 1996).
Stellar-mass black holes should also occur in isolation
 (remnant black holes, hereafter RBHs),
 although in the absence of a significant accretion fuel reservoir,
 their confirmation is expected to be a difficult task.
However,
 for those RBHs lying in high density environments such as molecular clouds
 and possessing small velocities relative to the surrounding gas,
 the rate of accretion directly from the ambient interstellar medium (ISM)
 may be sufficiently high for the resulting emission to be observable
 (\markcite{s1971}Shvartsman 1971).

More speculatively, black holes may also originate
 from processes other than conventional stellar evolution.
The discovery of microlensing events towards the Large Magellanic Cloud
 by the MACHO collaboration imply that a sizable fraction ($0.62^{+0.3}_{-0.2}$)
 of the dark matter in our Galactic halo may consist of compact objects (MACHOs)
 in the mass range $0.5^{+0.3}_{-0.2} \rm M_{\sun}$
 (\markcite{a1997}Alcock et al. 1997).
At present there is no consensus on the most likely identity of MACHOs,
 and a wide variety of proposals have been put forth.
While none are firmly excluded,
 the simplest sort of baryonic MACHOs, such as low-mass stars
 or collapsed objects
 resulting from ordinary stellar evolution
 all face various theoretical and observational challenges (e.g. \markcite{c1994}Carr 1994).
A serious possibility may then be that MACHOs are low-mass black holes (MACHOBHs)
 that were somehow created in the early universe;
 indeed, a few such scenarios have already been suggested
 (\markcite{y1997}Yokoyama 1997, \markcite{j97}Jedamzik 1997). 

Our objective here is to examine if accretion from interstellar gas
 onto RBHs or MACHOBHs in our Galaxy
 may give rise to observable radiation and aid in their identification.
Pregalactic black holes may also accrete from the surrounding medium
 and radiate at early epochs,
 but \markcite{c1979}Carr (1979) has shown that
 for the MACHOBH mass range of our interest (mass $\lesssim 1 \rm M_{\sun}$),
 no important constraints are imposed from the integrated background light,
 even if a high radiative efficiency is posited.
Thus, our consideration will be restricted to black holes in the solar neighborhood.

Most previous investigations of ISM-accreting black holes
 have employed spherical accretion models,
 such as that of Ipser \& Price (\markcite{ip1982}1982, \markcite{ip1983}1983) (hereafter IP).
However, there is a multitude of alternative spherical accretion models in the literature
 (for a review, see e.g. \markcite{c1996}Chakrabarti 1996), 
 and the emission spectra are sensitive to assumptions
 concerning the physical state of the accreting plasma and the radiation processes involved.
Since none of them have been subject to observational tests,
 the validity of the particular IP model is in question.
The advent of the advection dominated accretion flow (ADAF) model
 (\markcite{i1977}Ichimaru 1977, \markcite{rbbp1982}Rees et al. 1982,
 \markcite{ny1994}Narayan \& Yi 1994,
 \markcite{acklr1994}Abramowicz et al. 1995;
 see Narayan \markcite{n1996}1996, \markcite{n1997}1997 for reviews),
 being relevant for systems with relatively low accretion rates,
 has created a recent surge of interest in low-luminosity black holes.
With minimal assumptions and few free parameters,
 ADAFs self-consistently predict a stable, hot, two temperature structure
 that generates broad-band spectra from radio to gamma-rays,
 and have been successfully applied to a variety of low-luminosity objects.
The conditions for isolated, ISM-accreting black holes may also be appropriate for ADAFs,
 and in view of the fair amount of observational support,
 we adopt ADAF as the model of choice.
Note that \markcite{mq97}Mahadevan \& Quataert (1997)
 have also commented on the implications of ADAFs for isolated black holes with ISM accretion.

\section{Emission from RBHs}
\label{sec:rbh}

We first quote some standard results for accretion from a uniform medium.
 (\markcite{bh1944}Bondi \& Hoyle 1944, \markcite{fkr1992}Frank, King \& Raine 1992).
A black hole of mass $M$ moving with velocity $V$ with respect to the ambient medium
 captures gas within the accretion radius defined by
\begin{equation}
 \label{eq:rcap}
  r_{\rm cap} \sim \frac{2 GM}{V^2+{c_s}^2}\:,
\end{equation}
 where $G$ is the gravitational constant and $c_s$ is the medium's sound speed.
In the following we ignore $c_s$,
 which is justified so long as the gas is cold (as in molecular clouds)
 and does not lie in a hot phase of the ISM.
The accretion rate is
\begin{eqnarray}
 \label{eq:dotm}
 \dot{M} &\sim& \pi r_{\rm cap}^{2}\rho_{\rm gas}V \nonumber \\
     &\approx& 7.4\times 10^{13}\;{\rm g\;s^{-1}} \left(\frac{M}{M_{\sun}}\right)^{2}
          \left(\frac{n_{\rm gas}}{10^{2}\;\rm cm^{-3}}\right)
          \left(\frac{V}{10\;\rm km\;s^{-1}}\right)^{-3}\; \nonumber \\
     &\approx& 5.3\times 10^{-4} \dot{M}_{\rm Edd} \left(\frac{M}{M_{\sun}}\right)
          \left(\frac{n_{\rm gas}}{10^{2}\;\rm cm^{-3}}\right)
          \left(\frac{V}{10\;\rm km\;s^{-1}}\right)^{-3}\;,
\end{eqnarray}
 where $\rho_{\rm gas}$ and $n_{\rm gas}$ are respectively
 the mass and number density of the medium,
 and ${M}_{\rm Edd}= 1.39\times 10^{17}\;(M/M_{\sun})\;{\rm g\;s^{-1}}$
 is the Eddington accretion rate with unit efficiency.

Whether the gas then accretes onto the black hole more or less spherically
 or through a disk-like configuration
 depends on the amount of angular momentum available in the ISM.
One possible source is ISM turbulence,
 for which the turbulent velocity $v_{\rm turb}$ has been measured to correlate
 with scale $r$ as $v_{\rm turb} \sim 1.1\;{(r/{\rm 1pc})}^{0.38}{\rm km\;s^{-1}}$,
 albeit with large observational errors
 (\markcite{l1981}Larson 1981, \markcite{fp1990}Falgarone \& Phillips 1990).
Adopting this relation,
 the criterion that a rotating accretion flow can extend out to radius $r_o$
 with specific angular momentum $f_l$ times the Keplerian value
 is that the black hole velocity satisfies
 $V \lesssim 52\;{\left({r_g / r_o {f_l}^{2}}\right)}^{0.18} (M/M_{\sun})^{0.14}{\rm km\;s^{-1}}$,
 where $r_g=2GM/c^2$ is the Schwarzschild radius (Ipser \& Price \markcite{ip1977} 1977).
Inhomogeneties in the medium (which is ubiquitous in the ISM)
 may also lead to accretion with angular momentum (\markcite{ip1977}Ipser \& Price 1977),
 although there are conflicting opinions on this matter
 (\markcite{dp1980}Davies \& Pringle 1980, \markcite{abm1987}Anzer, Borner \& Monaghan 1987).
ADAFs actually allow self-consistent solutions
 even for sub-Keplerian outer boundary conditions (see below),
 and we believe that they represent the most legitimate mode of accretion from the ISM.

The principal assumptions of the ADAF paradigm are:
 1) $\alpha$ viscosity prescription (\markcite{ss1973}Shakura \& Sunyaev 1973);
 2) viscous heat primarily channeled into protons;
 3) proton-electron energy transfer mediated by Coulomb collisions; and
 4) existence of magnetic fields near equipartition.
Models have been constructed for black holes of $10 \rm M_{\sun}$, typical of RBHs,
 accreting at different rates
 (corresponding to different combinations of $n_{\rm gas}$ and $V$, Eq. [\ref{eq:dotm}])
 in the ISM.
We exploit global solutions of the dynamical equations,
 which include the important effect of energy advection and compressional heating for electrons
 (\markcite{nmkk1996}Nakamura et al. 1996, \markcite{nkmk1997a}1997a, 
 \markcite{emn1997}Esin et al. 1997).
The prescription for obtaining the solutions and computing the emission spectra
 are identical to that described in \markcite{mmk1997}Manmoto, Kusunose \& Mineshige (1997),
 except here the angular momentum has been set to half the Keplerian value ($f_l=0.5$)
 at $r_o=1000r_g$
\footnote{Strictly speaking, this outer boundary condition conforms
 to the above ISM turbulence relation
 only for objects with $V \lesssim 19\;(M/M_{\sun})^{0.14}{\rm km\;s^{-1}}$.
Nonetheless, in view of the large observational uncertainties,
 along with the fact that the bulk of the ADAF emission originates from $\lesssim 100 r_g$,
 the models should also be suitable for even higher velocity black holes.}.
The free parameters in the model are chosen to be
 $\alpha = 0.1$ for the dimensionless viscosity, and
 $\beta = 0.5$ for the ratio of gas pressure to total pressure.

The results are shown in Figure 1.
In contrast to the IP spherical accretion model
 or the standard thin disk model of \markcite{ss1973}Shakura \& Sunyaev (1973),
 the spectrum arising from the hot, optically thin ADAF
 is very broad band and consists of roughly three peaks.
Going from low to high frequencies, the processes responsible for each are
 synchrotron radiation (radio to optical),
 single Compton upscattering (UV to soft X-ray),
	and bremsstrahlung augmented by multiple Compton upscattering
 (hard X-ray to soft gamma-ray).
Since hard X-rays are little affected by interstellar absorption,
 the occurrence of the high-energy component
 is a crucial point for the observability of RBHs.

\begin{figure}[htb]
 \label{figrbh}
 \plotone{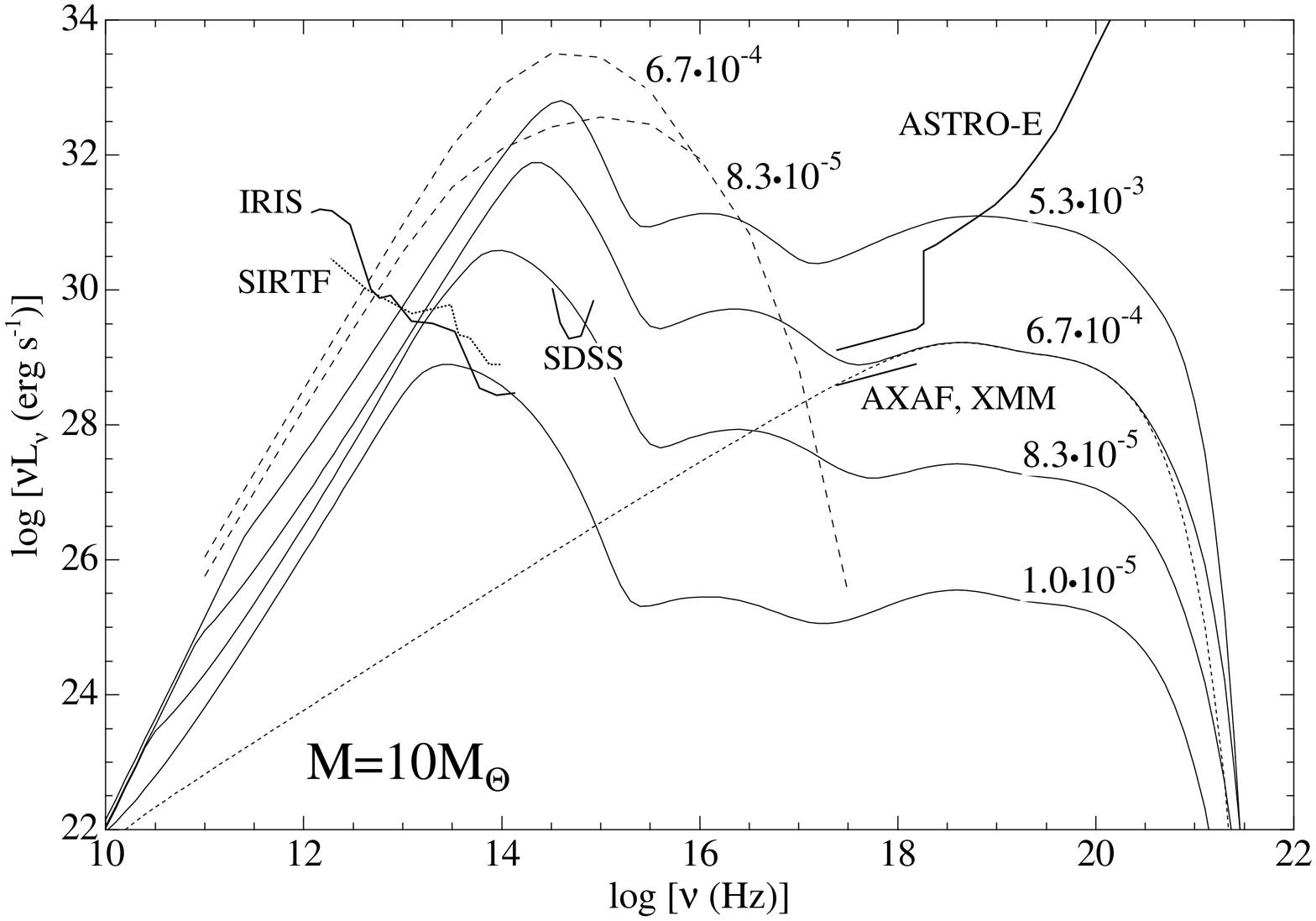}
\caption[fig1.eps]
{
The luminosity spectra of RBHs in the ADAF model (solid lines)
 with $M=10\;\rm M_{\sun}$ and four different values of $\dot{M}$
 (marked on each curve in Eddington units),
 which correspond to black hole velocities
 $V=10, 20, 40, 80\;\rm km\;s^{-1}$ for $n_{\rm gas}=10^{2}\;\rm cm^{-3}$,
 or $V=2.2, 4.3, 8.6, 17.2\;\rm km\;s^{-1}$ for $n_{\rm gas}=1\;\rm cm^{-3}$.
The bremsstrahlung component alone is exemplified by the dotted line 
 in one case ($\dot{M}=6.7 \times 10^{-4} \dot{M}_{\rm Edd}$).
To be compared are the Ipser-Price model spectra (dashes lines) for two $\dot{M}$ values.
The typical point source detection limits
 for various observational facilities mentioned in the text have also been indicated,
 assuming a distance of 400pc (e.g. Orion).
}
\vspace{1cm}
\end{figure}

With these results in hand, we may assess the number of RBHs
 that are potentially detectable by near future instruments.
As the nearest example of a high density region of the ISM, we take the Orion molecular cloud,
 with a size of $\sim 50$ pc, density $n_{\rm gas}\sim 10^{2}\;\rm cm^{-3}$
 and distance $\sim 400$ pc (\markcite{lsm1993}Lada, Strom, \& Myers 1993).
RBHs in this cloud would generate X-ray luminosities
 in excess of $\sim 10^{29}\;\rm erg\;s^{-1}$ at $\sim 4$ keV
 if they move at $V \lesssim 20\;\rm km\;s^{-1}$ (Figure 1);
 this translates into fluxes $\sim 5\times 10^{-15}\;\rm erg\;s^{-1}\;cm^{-2}$,
 which are above or near the expected detection limits of
 {\em AXAF}, {\em XMM} and {\em ASTRO-E}.
Assuming RBHs obey a Maxwellian velocity distribution,
 the fraction of RBHs moving slower than $V$ is
\begin{equation} 
 \label{eq:frbh}
 f(<V) =  \frac{1}{(2\pi)^{3/2}\sigma^{3}}
 \int_{0}^{V} \exp\left(-\frac{v^{2}}{2\sigma^{2}}\right) 4\pi v^{2} dv\;.
\end{equation}
The present velocity dispersion $\sigma$ of RBHs is an unknown quantity,
 and depends on the their uncertain initial velocities and positions
 (\markcite{cp1993}Campana \& Pardi 1993).
Here we simply take $\sigma =30\;\rm km\;s^{-1}$,
 the velocity dispersion of population I stars,
 which results in $f(<20\;\rm km\;s^{-1}) =0.07$. 
The average number density of $10 \rm M_{\sun}$ RBHs in the solar neighborhood
 has been estimated in a simplified manner by \markcite{st1983}Shapiro \& Teukolsky (1983)
 to be $8\times 10^{-4}\rm \; pc^{-3}$.
With this value, the Orion molecular cloud would harbor a total of $\sim$100 RBHs,
 of which $\sim 7$ could be detected by the aforementioned X-ray observatories.
More detailed studies of RBH formation and evolution
 have been conducted by \markcite{cp1993}Campana \& Pardi (1993),
 and if their somewhat model-dependent results are used,
 the expected number becomes $\lesssim 6$.

The most slowly moving RBHs may also be observable
 even if they do not reside in molecular clouds.
RBHs with $V =4\;\rm km\;s^{-1}$
 in the average ISM density of $n_{\rm gas}=1\;\rm cm^{-3}$ can radiate as much as
 those with $V = 20 \;\rm km\;s^{-1}$ and $n_{\rm gas}=10^{2}\;\rm cm^{-3}$
 (Eq. [\ref{eq:dotm}]).
Thus the same X-ray instruments as above
 are capable of detecting RBHs in the general ISM within 400pc of the Sun.
With $f(<4\;\rm km\;s^{-1}) = 6.3\times 10^{-4}$ for $\sigma = 30\;\rm km\;s^{-1}$
 and the thickness of Galactic disk being $\sim 100$ pc,
 we obtain in total $\sim 25$ such RBHs.

Observations at lower frequencies should also play a role in the identification of RBHs,
 as their power output reaches a maximum in the infrared-optical region. 
The expected mid-IR flux from RBHs at the distance of Orion are
 $\sim$ 0.1 mJy at $\sim 10\;\mu m$ for $V \lesssim 50\;\rm km\;s^{-1}$,
 above the typical detection thresholds of telescopes
 such as {\em IRIS} or {\em SIRTF}.
Wide-field surveys by these facilities planned to hunt for brown dwarfs
 could uncover a sizable number of RBHs as a byproduct.
Heavy extinction may completely obscure the optical to UV emission
 from RBHs within molecular clouds,
 and should reduce their near-IR and soft X-ray fluxes as well,
 although these wavebands may still be accessible for RBHs outside dense parts of the ISM.

To discriminate RBHs from other IR or optical sources (e.g. young stellar objects),
 one calls for more information than narrow band spectra.
Several effects could be of help:
 variability on various timescales from $\lesssim 1$ day to $\sim 1$ yr,
 possibly arising from fluctuations in accretion rate caused by ISM inhomogeneities;
 locational distribution, expected to be more or less uniform throughout our neighborhood,
 as opposed to young stars preferentially inhabiting molecular cloud cores
 with small filling factors ($<0.01$);
 and proper motions which may be detectable for the more closer ones.
Yet the most powerful method of identifying RBHs should be combining multiwavelength data.
A preferred strategy may be to search first for candidate hard X-ray sources,
 where there is the least worry over interstellar absorption or confusing objects,
 and then conduct radio, IR and optical follow-up observations
 to see if the anticipated spectral distribution is revealed;
 this should prove effective in separating RBHs from other X-ray sources.

We point out here one theoretical uncertainty regarding the low frequency emission from ADAFs.
Our models assume the particles are always thermal,
 but Mahadevan \& Quataert (1997) have recently shown that
 ADAFs at accretion rates $\dot{M} \lesssim 0.4 {\alpha}^2 \dot{M}_{\rm Edd}$
 contain regions where the electrons do not thermalize,
 their energy distribution being determined by the relevant heating and cooling processes.
At $\dot{M} \lesssim 10^{-3} {\alpha}^2 \dot{M}_{\rm Edd}$
 this applies throughout the flow,
 in which case a nonthermal distribution
 with a reduced fraction of particles in the high-energy tail may be realized,
 leading to large decreases in the synchrotron and Compton emission components.
The consequences should not be so pronounced, however,
 for emission components that are caused mainly by electrons with the plasma's mean energy,
 such as bremsstrahlung (indicated separately for one case in Figure 1);
 thus the hard X-ray fluxes, critical to the present discussion,
 are not greatly modified by this effect.

ADAFs may also emit high energy gamma-rays
 at luminosities comparable to the bremsstrahlung component
 as a result of pion production in collisions between hot protons
 (\markcite{mnk1997}Mahadevan, Narayan \& Krolik 1997).
Despite the known association of some unidentified {\em EGRET} sources
 with high density regions of the ISM (e.g. \markcite{yr1997}Yadigaroglu \& Romani 1997),
 the gamma-ray luminosities expected from RBHs are probably too low
 to account for any of these sources.
Nevertheless, next generation instruments like {\em GLAST}
 may be sensitive enough to detect gamma-ray emission from the nearest RBHs.

The above conclusions are in marked contrast to previous studies
 (\markcite{m1985}McDowell 1985, \markcite{cp1993}Campana \& Pardi 1993,
 \markcite{hk1996}Heckler \& Kolb 1996) with the IP model,
 the results of which are overlayed in Figure 1 for comparison.
The IP spherical accretion model is based on the ad hoc assumption
 of equipartition between gravitational, kinetic and magnetic energies,
 and possesses a single, high temperature,
 which results in synchrotron emission chiefly in the IR to soft X-ray range.
At the pertinent accretion rates the hard X-ray flux from Comptonization in this model
 (\markcite{ip1983}Ipser \& Price 1983) is insignificant,
 and as UV to soft X-rays suffer severe interstellar attenuation,
 visibility is practically limited to the IR-optical range.
As noted above, IR-optical spectral data is insufficient
 for unambiguously distinguishing RBHs from other sources in the same waveband
 which are far more numerous,
 and one must rely on difficult measurements of variability or proper motion.
RBHs accreting via ADAFs should be much easier to distinguish by their high energy emission.

\markcite{hk1996}Heckler \& Kolb (1996) have recently suggested that
 the color-based source selection scheme and spectral measurements
 intended for QSO searches in the {\em Sloan Digital Sky Survey (SDSS)}
 may also be advantageous for discerning isolated black holes that radiate a la Ipser \& Price.
Although ADAFs predict optical colors notably different from the IP model,
 and fluxes which are lower by about two orders of magnitude
 (and perhaps even lower in the case of a nonthermal electron distribution),
 Figure 1 suggests that the X-ray detectable RBHs
 are also visible to the {\em SDSS}, if extinction is not serious.
With its exceptionally large field of view, the {\em SDSS} could still be useful
 if its results are utilized in conjunction with X-ray information.

\section{Emission from MACHOBHs}
\label{sec:MBH}

As remarked in the introduction,
 there is the intriguing possibility that a large number of low-mass black holes
 comprises the dark halo of our Galaxy and causes the microlensing events.
Here we choose a MACHOBH mass of $0.5 \rm M_{\sun}$,
 the average value implied by microlensing event durations.
The emergent spectra in the ADAF model have been calculated for such MACHOBHs
 with the same procedure as for RBHs and are shown in Figure 2.

\begin{figure}[htb]
 \label{figmabh}
 \plotone{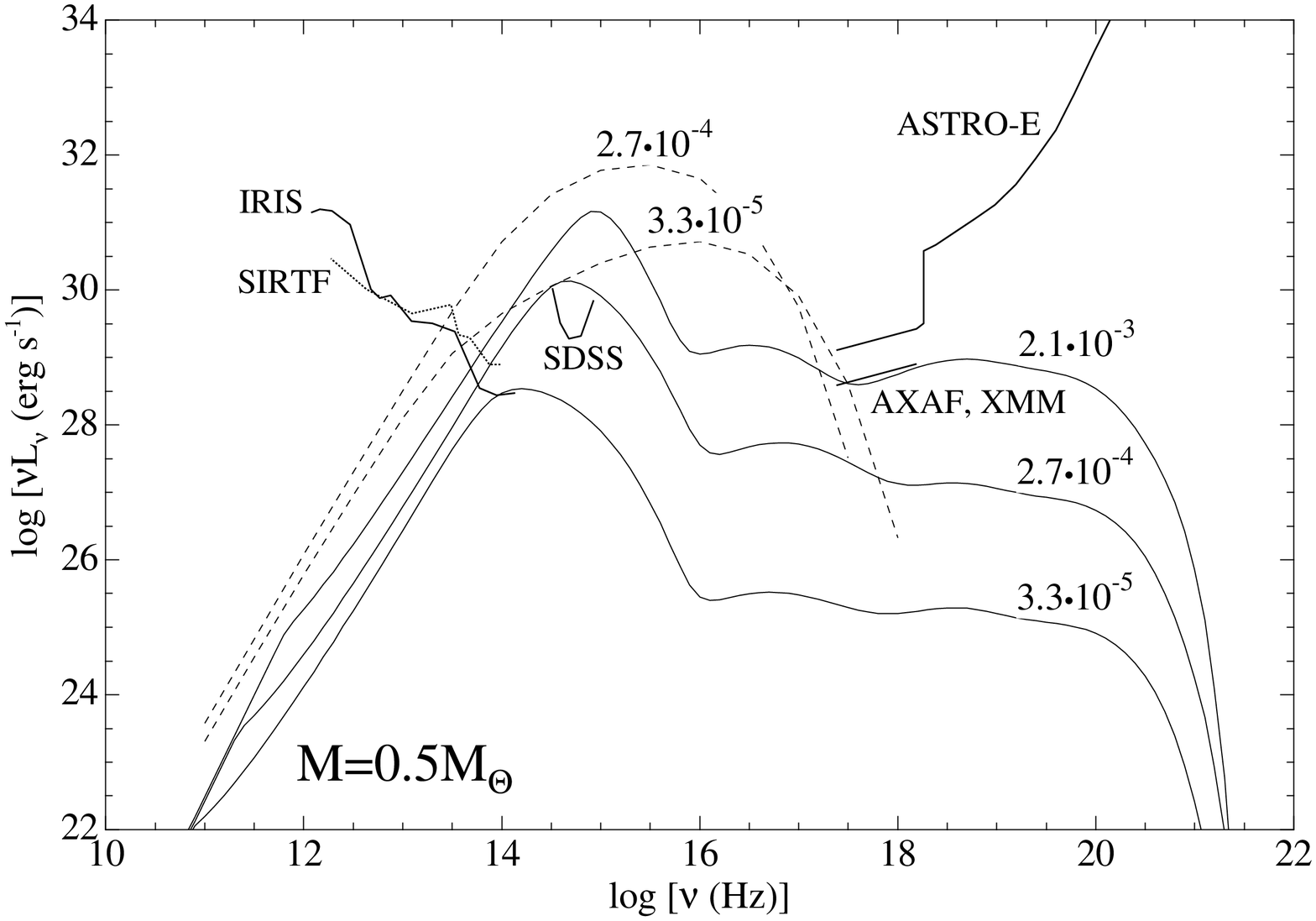}
\caption[fig2.eps]
{
The luminosity spectra of MACHOBHs in the ADAF model (solid lines)
 with $M=0.5\;\rm M_{\sun}$ and three different values of $\dot{M}$,
 corresponding to $V=5, 10, 20\;\rm km\;s^{-1}$ for $n_{\rm gas}=10^{2}\;\rm cm^{-3}$.
Also shown are the Ipser-Price model spectra (dashes lines) for two $\dot{M}$ values.
Otherwise similar to Fig.1.
}
\vspace{1cm}
\end{figure}

The smaller mass and lower accretion rate (Eq. [\ref{eq:dotm}])
 compared to RBHs make the overall luminosities of MACHOBHs substantially lower.
For a medium of $n_{\rm gas}=10^2\rm\; cm^{-3}$, $400$ pc away,
 observability demands velocities
 $V\lesssim 20\;\rm km\;s^{-1}$ for the near-IR band,
 and $V\lesssim 4\;\rm km\;s^{-1}$ for X-rays.
The fraction of MACHOBHs whose velocities are smaller than $V$ is given by
\begin{eqnarray} 
 \label{eq:frac}
  f_{\rm MA}(<V)& = & \frac{1}{(2\pi)^{3/2}\sigma_{\rm MA}^{3}}
  \int_{0}^{V}v^{2}dv \int_{0}^{2\pi}d\phi
 \int_{0}^{\pi}d\theta\sin\theta \exp\left
 (-\frac{v^2+V_{\rm rot}^2+2vV_{\rm rot}\cos\theta}
 {2\sigma_{\rm MA}^2}\right)
 \nonumber \\
  & = &
 \sqrt{\frac{2}{\pi}}\frac{1}{V_{\rm rot}\sigma_{\rm MA}}
 \exp\left(-\frac{V_{\rm rot}^2}{2\sigma_{\rm MA}^2}\right)  
 \int_{0}^{V}dv v\sinh\left(\frac{vV_{\rm rot}}{\sigma_{\rm MA}^2}\right)
 \exp\left(-\frac{v^2}{2\sigma_{\rm MA}^2}\right) \;.
\end{eqnarray}
Here $\sigma_{\rm MA}$ is the MACHO velocity dispersion,
 and $V_{\rm rot}$ is the Galactic rotation velocity in the solar neighborhood
 with the direction of rotation being $(\theta,\phi)=(0,0)$.
We take $\sigma_{\rm MA}=155\;\rm km\;s^{-1}$ and $V_{\rm rot}=220\;\rm km\;s^{-1}$,
 and assume the density of dark matter in the solar neighborhood
 to be $0.0079 \;\rm M_{\sun}\;pc^{-3}$ (\markcite{a1996}Alcock et al. 1997),
 which gives a MACHOBH number density of $0.016\;\rm pc^{-3}$. 
As $f_{\rm MA}(<20\;\rm km\;s^{-1}) = 2\times 10^{-4}$,
 the number of near-IR observable objects in Orion is only $\sim 0.4$,
 while the chances of an X-ray detection are hopelessly small.
For MACHOBHs in the general ISM of $n_{\rm gas}=1\rm\; cm^{-3}$
 the detection requirements are
 $V \lesssim 4\; \rm km\; s^{-1}$ (near-IR) and $V \lesssim 0.9\; \rm km\; s^{-1}$ (X-ray)
 at distances $\lesssim 400$ pc;
 $f_{\rm MA}(<4\;\rm km\;s^{-1}) = 2\times 10^{-6}$
 then results in $\sim$ 2 near-IR MACHOBHs, with the X-ray numbers again being miniscule.

MACHOBH spectra in the IP spherical accretion model are also shown in Figure 2.
Were the model applicable, its greater luminosity at optical wavelengths
 may raise the number of detections to $\sim 10$.
As remarked in the previous section, however,
 discriminating isolated, accreting black holes
 from other sources based solely on IR-optical observations
 is akin to searching for needles in a haystack.
Applying either the ADAF model or the IP model,
 we must conclude that identification of MACHOBHs by their accretion-driven emission
 is extremely difficult if not impossible.

\section{Conclusions}
Implementing the advection dominated accretion flow model,
 we have considered accretion from the interstellar medium
 and the resultant emission for black holes of two types,
 those originating as stellar remnants and those constituting MACHOs.
A fair number of stellar remnant black holes
 may be observed by X-ray, infrared and optical telescopes in the near future,
 which would provide strong proof of their existence.
MACHO black holes, due to their smaller mass and larger velocity on average,
 are probably too underluminous for clear identification;
 gravitational radiation from binary MACHOBH mergers
 may be the only means to see them directly in the foreseeable future
 (\markcite{nstt1997b}Nakamura et al. 1997b).

We may ask one final question: could accretion-induced radiation
 be of any significance for the case of neutron star (NS) or white dwarf (WD) MACHOs?
Unlike black holes,
 radiation from accretion onto a NS or a WD
 is dominated by thermal emission from their surfaces where the radiative efficiency is high,
 and ambiguities in the model of accretion is of no serious concern.
Such objects accreting from the ISM
 should indeed be capable of producing observable radiation if they are slow enough.
The main difficulty we face here is misidentification with objects evolved from disk stars:
 simple estimates demonstrate that NS or WD MACHOs
 which meet the criterion for detectability
 are outnumbered by objects of disk origin with similar properties.
Thus identifying them by their accretion-induced emission is also problematic.

\acknowledgements

We thank
 H. Hanami, F. Iwamuro, Y. Kan-ya, S. Mineshige, T. Ohashi, S. Oya, M. Sakano and N. Yamasaki
 for helpful discussions.
We are also grateful to an anonymous referee for several suggestions
 which improved the paper.
This work was supported in part by the JSPS Research Fellowship for Young Scientists,
 and Grant-in-Aid of the Ministry of Education, Culture, and Sports No.09640351.




\end{document}